\documentclass[prl,reprint,a4paper,superscriptaddress,citeautoscript,floatfix,nofootinbib]{revtex4-1}

\pdfoutput=1
\usepackage[charter]{mathdesign}
\usepackage{amsmath}
\usepackage{siunitx}
\usepackage[pdftex]{graphicx}
\usepackage{microtype}
\usepackage{bm}

\usepackage[svgnames]{xcolor}
\usepackage[
	colorlinks=True,linkcolor=DarkRed,citecolor=ForestGreen,urlcolor=MediumBlue,
	pdfstartview=FitH,bookmarks=False,pdfpagemode=UseNone
]{hyperref}

\newcommand{\cdensityunit}[1]{\textrm{$#1\times 10^{12}~\textrm{cm}^{-2}$}}
\newcommand{\alphaunit}[1]{\textrm{$#1\times 10^{-12}~\textrm{eV~m}$}}
\newcommand{\mobunit}{~\textrm{cm$^2$V$^{-1}$s$^{-1}$}}
\newcommand{\Ref}[1]{(\ref{#1})}

\begin{document}

\title{\boldmath Large modulation of the Shubnikov-de Haas oscillations by the Rashba interaction at the LaAlO$_{3}$/SrTiO$_{3}$ interface}

\author{A. F{\^e}te}
\author{S. Gariglio}
\author{C. Berthod}
\author{D. Li}
\author{D. Stornaiuolo}
\affiliation{D{\'e}partement de Physique de la Mati{\`e}re Condens{\'e}e, Universit{\'e} de Gen{\`e}ve, 24 Quai Ernest-Ansermet, 1211 Gen{\`e}ve 4, Suisse}
\author{M. Gabay}
\affiliation{Laboratoire de Physique des Solides, Bat. 510, Universit\'e Paris-Sud 11,
Centre d'Orsay, 91405 Orsay Cedex, France}
\author{J.-M. Triscone}
\affiliation{D{\'e}partement de Physique de la Mati{\`e}re Condens{\'e}e, Universit{\'e} de Gen{\`e}ve, 24 Quai Ernest-Ansermet, 1211 Gen{\`e}ve 4, Suisse}

\begin{abstract}
We investigate the 2-dimensional Fermi surface of
high-mobility LaAlO$_3$/SrTiO$_3$ interfaces using Shubnikov-de Haas
oscillations. Our analysis of the oscillation pattern underscores the key role played by the Rashba spin-orbit interaction brought about by the breaking of inversion symmetry, as well as the dominant contribution of the heavy $d_{xz}$/$d_{yz}$ orbitals on electrical  transport. We furthermore bring into light the  complex evolution of the oscillations with the carrier density, which is tuned by the field effect.
\end{abstract}
\maketitle

The conducting interface between the two band insulators LaAlO$_3$ (LAO) and SrTiO$_3$ (STO) has drawn a lot of attention as it presents a variety of exciting properties, among them superconductivity and a large spin-orbit coupling, both being tunable by an electric field \cite{Zubko2011}. As the 2DEG lies on the STO side, the conduction band of the system is dominated by the Ti $3d$-$t_{2g}$ orbitals as for bulk STO. However, at the interface, quantum confinement spectacularly alters the orbital
ordering of the energy levels, as observed by X-ray spectroscopy
\cite{Salluzzo2009}: for a given sub-band index, the states with predominantly
$d_{xy}$ symmetry have, on average, a lower energy than states derived from the
$d_{xz}$/$d_{yz}$ orbitals. Currently, experimental and theoretical estimates of
the out-of-plane extent of the 2DEG vary from a few monolayers \cite{Sing2009,
Cancellieri2013, Delugas} to 10~nm \cite{Reyren2009, Copie2009, Khalsa2012} and,
consequently, the number and precise energy arrangement of these sub-bands is
still an open question. The asymmetric confining potential also brings about a
breaking of inversion symmetry: its effect is to spin-split the electronic bands (Rashba effect) \cite{Zhong2013, Kim2013, Khalsa2013} with important consequences on the magnetotransport of the system 
 \cite{CavigliaWAL,
BenShalom,AlexAMRO}. 

In this letter, we report the observation and analysis of Shubnikov-de Haas (SdH) oscillations in high-mobility and low carrier density ($\approx \SI{e12}{\per \square \centi \meter}$) interfaces. Quantum oscillations show two frequencies that we contend are due to the splitting of an electronic band induced by the Rashba spin-orbit interaction (SOi). The estimated SOi energy is comparable to the Fermi energy ($E_\textrm{F}$), defining an unusual regime when compared to semiconductor 2DEG. Electric field effect experiments also reveal that the evolution of the Landau levels (LLs) that is observed as one changes the carrier density is singular.

The LaAlO$_3$ layers were grown by pulsed laser deposition at \SI{650}{\celsius}, a lower
temperature than for standard interfaces \cite{CavigliaSdH}. Hall bars for DC transport
measurements were patterned and field-effect devices were realized using the STO
single crystal substrate as the gate dielectric (see supplementary data). Magnetotransport measurements were performed in a dilution
refrigerator equipped with a 8~T superconducting magnet.

\begin{figure*}
\centering\includegraphics[width=1.4\columnwidth]{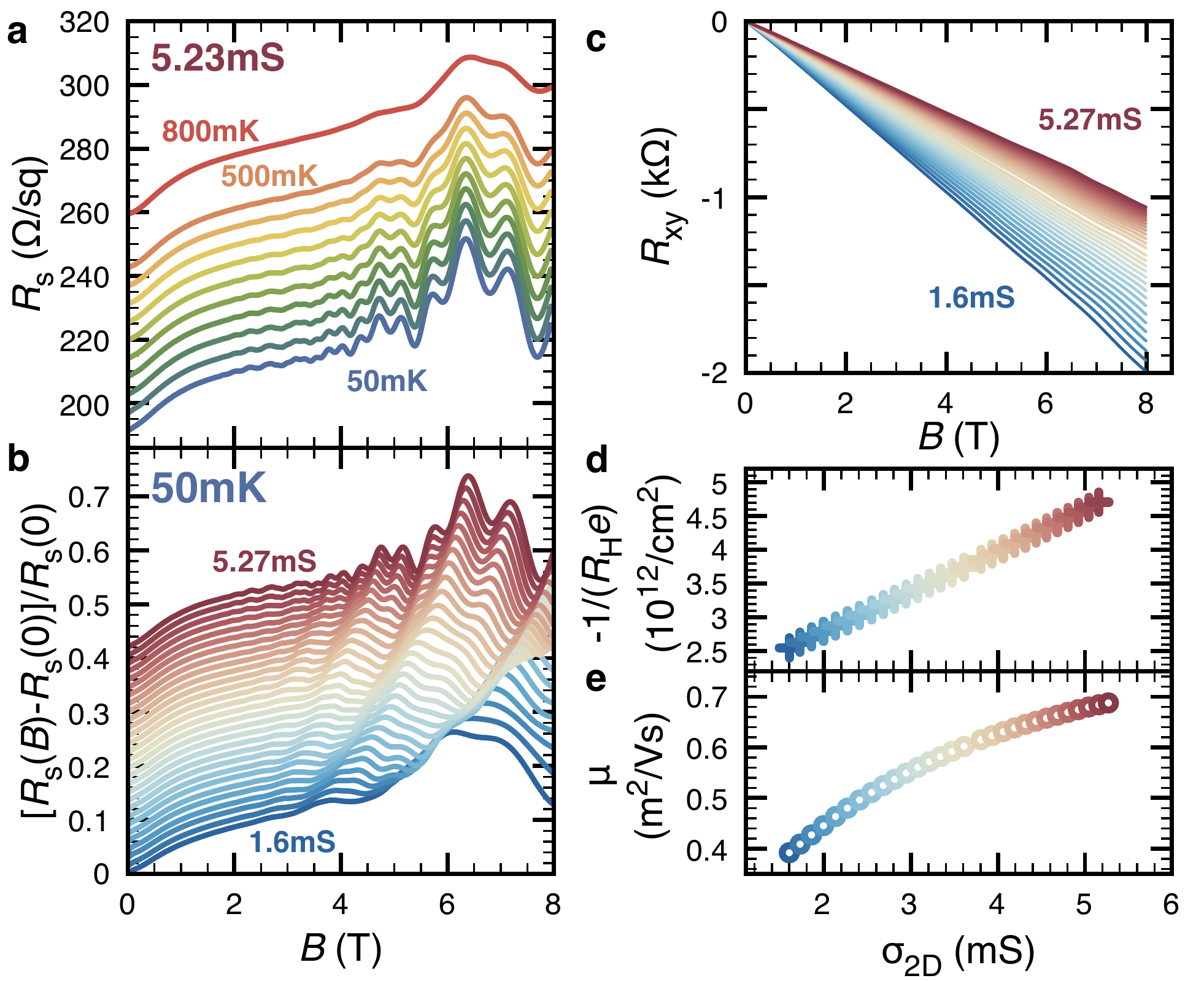}
\caption{\label{Rawdata}
Transport properties in a magnetic field. (a) Temperature
evolution of the sheet resistance ($R_s$) versus magnetic field for a doping
level corresponding to a sheet conductance of 5.23~mS and a mobility of
$\mu\approx7000\mobunit$ at 50~mK. Curves are offset for clarity. In this paper,
the sheet conductance at 50~mK and \SI{0}{\tesla} ($\sigma_{\textrm{2D}}$) is used as a reference
for the doping level. (b) $\left[R_s(B)-R_s(0)\right]/R_s(0)$ for
different dopings, illustrating the evolution of the SdH oscillations with gate
voltage. Curves are offset for clarity. (c) Hall resistance versus
magnetic field at 50~mK for different dopings. (d) Inverse Hall
coefficient and (e) the corresponding Hall mobility at 50~mK versus $\sigma_{\textrm{2D}}$.
}

\end{figure*}

Figure~\ref{Rawdata}a displays a set of sheet resistance versus magnetic field ($B$)
curves for temperatures ranging from 800~mK to 50~mK. As can be seen, the
magnitude of the Shubnikov-de Haas oscillations increases markedly as the
temperature is lowered. At 50~mK and in high magnetic field, the amplitude of
SdH oscillations is about 10--15\% of the sheet resistance value. To change the
carrier density, we apply a back-gate voltage \footnote{In the rest of the paper we use the sheet conductance at zero magnetic field and 50~mK ($\sigma_\textrm{2D}$) rather than the gate voltage ($V_\textrm{g}$) to define the state of the system.}. Fig.~\ref{Rawdata}c shows that upon carrier density tuning
the transverse resistance $R_{xy}$ varies linearly with magnetic field. From Fig.~\ref{Rawdata}d, we see that ramping the gate voltage ($V_g$)
up to large positive values leads to an increase of the inverse Hall coefficient.
Analysing the Hall signal using a single-band model, we extract a carrier
density at \SI{50}{\milli \kelvin} that increases from 2.5 to \cdensityunit{4.8} as $V_g$ is swept from 79
to 107~V, \textit{i.e.} as the sheet conductance ($\sigma_\textrm{2D}$) at \SI{0}{\tesla} increases from 1.6 to 5.27~mS. Concomitantly with this variation of the electron density, the electron mobility $\mu$
evolves from 3900 to 6900\mobunit, as shown in Fig.~\ref{Rawdata}e \cite{Bell2009a, AlexAMRO,Joshua2012}. We note that these samples exhibit $n_\textrm{2D}$ ($\mu$) that are smaller (larger) than standard samples. Moreover, the modulation of the carrier density and mobility by electric field effect does not induce a transition from linear to non-linear Hall effect.

As can be seen in Fig.~\ref{Rawdata}b, the changes in
electron mobility and density strongly modify the structure of the SdH
oscillations with a clear change in both the peak position and the period of
the oscillations.

In order to proceed with the analysis of the SdH data presented in
Fig.~\ref{Rawdata}, we subtracted the background: 
	\begin{equation}
		\Delta\sigma(B)=\frac{R_{s}(B)}{[R_{s}(B)]^2+[R_{xy}(B)]^2}
		-\sigma_0(B)
	\end{equation}
with $R_{s}(B)$ and $R_{xy}(B)$ the measured longitudinal and transverse
resistances, respectively, and $\sigma_0(B)$ a non-oscillating polynomial background. Examples of the resulting curves can be found in Fig.~\ref{Fig4}.

Looking at the SdH oscillations, we note (at least) two frequencies modulating the conductance. Hence, we first analyse the data considering a model with two parabolic bands for which the magnetoconductance can be calculated using the Lifshitz-Kosevich (LK) formula \cite{LK}. We fit the data for the largest conductance introducing an arbitrary phase for each frequency.

\begin{figure*}
\centering\includegraphics[width=4.2in]{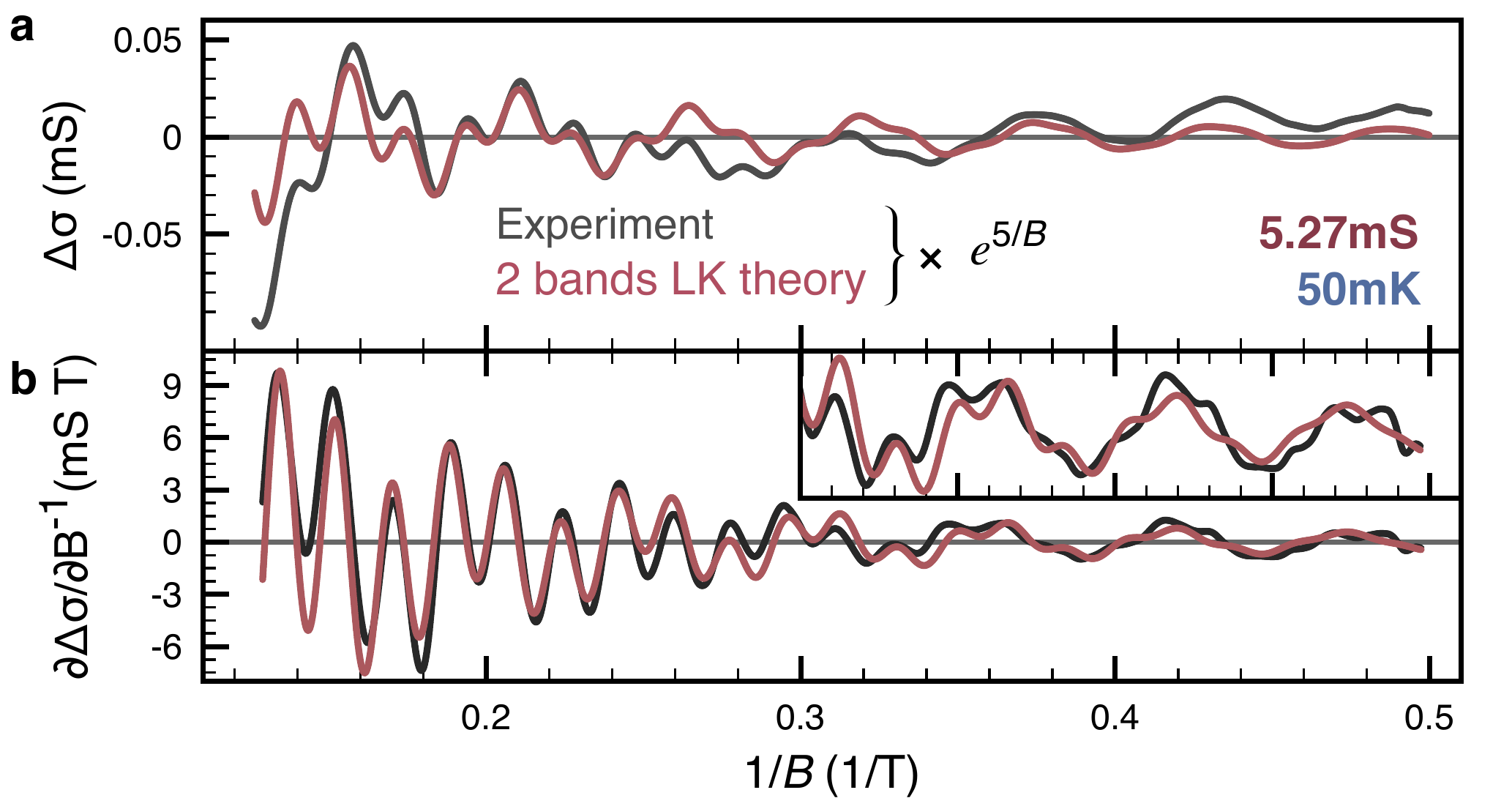}
\caption{\label{MBmodel}
(a) Comparison between $\Delta\sigma$ versus $1/B$ calculated
within the two-band model (red) and the experimental data (black) for the doping
with the highest conductance (5.27~mS) and at \SI{50}{\milli \kelvin}. The exponential factor $e^{5/B}$ is
used to magnify the low-field region. (b) Derivative with
respect to $B^{-1}$ of the curves presented in (a). (inset) Close-up on the low-field region of (b).}
\end{figure*}

As can be seen from Fig.~\ref{MBmodel}a, a good fit to the data can be obtained using the two frequencies 18~T and 55.9~T \footnote{We would like to emphasize that at our base temperature ($k_\protect \textrm {B} T \approx 10^{-3}$~meV) the field dependence of the amplitude of the oscillations is mainly controlled by the Dingle term that compares the strength of the disorder to the cyclotron gap. Hence, the fitting shown in Fig.~\ref {MBmodel} is only sensitive to the product of $m^*$ and the Dingle temperature.}. Panel b displays the derivatives of the
theoretical and experimental curves allowing the positions of the maxima and
minima to be compared. Considering the Onsager relation
with a spin degeneracy of $\nu_s=2$ and a valley degeneracy of $\nu_v=1$, we
find the carrier densities for the two bands to be 0.87 and \cdensityunit{2.7},
which yield a total carrier concentration of $\sim\cdensityunit{3.6}$.

In the LK formalism, the temperature evolution of the oscillations can be directly related to the effective mass of the oscillating carriers. We extracted the high and low frequency (HF and LF)
parts of the SdH oscillations shown in Fig.~\ref{MBmodel}a. Selecting 17 and 4
extrema for the HF and LF, respectively, good agreement between theory and experiment is obtained by choosing an
effective mass of $2.7 m_e$ for the HF and $1.25 m_e$ for the LF (see supplementary data).

With the information extracted from this analysis, the electronic structure of our
two-band model can be reconstructed and the splitting at the Fermi level between
the heavy and the lighter bands determined:
	\begin{equation}\label{Esplitting}
		\Delta E  = \left| E_1(\bar{k}_{\textrm{F}})-E_2(\bar{k}_{\textrm{F}})\right|  \textrm{,  }
		E_i(k)  =\frac{\hbar^2}{2 m^*_i}\left(k^2-k_{\textrm{F},i}^2\right).
	\end{equation}
$k_{\textrm{F},i}$ is the Fermi momentum in the $i$-th band obtained from
the area $A_i=\pi k_{\textrm{F},i}^2$ calculated using the Onsager relation, and
$\bar{k}_{\textrm{F}} = (k_{\textrm{F},1}+k_{\textrm{F},2})/2$. We find $\Delta
E \approx 2.45$~meV.

The band structure obtained in the above two carrier model predicts a heavy band with a higher binding energy than the light one. This is in apparent contradiction with the well documented observation of orbital reconstruction at the LAO/STO interface \cite{Salluzzo2009}.

Another possible scenario is that the calculated band splitting $\Delta E$ is in reality the Rashba spin-orbit splitting estimated for LAO/STO heterostructures \cite{CavigliaWAL, BenShalom}. In what follows, we hence consider a model consisting of a single parabolic band split by the Rashba spin-orbit interaction.

In a Rashba scenario, the SOi splits
the LLs of a single band into two families ($\pm$) of irregularly-spaced levels.
These energy levels are labeled by an integer $N\geqslant 0$ and read, for an
isotropic Fermi surface and a $k$-linear splitting \cite{Winkler}:
	\begin{equation}\label{RZLLs}
		E_{N=0}  =E_c/2-E_Z \textrm{, }
		E_{N>0}^\pm  =N E_c\mp\sqrt{(E_c/2-E_Z)^2+N E_\alpha^2}\ .
	\end{equation}
$E_c=\hbar\omega_c^{*}$, $E_Z=(g^*/2)\mu_\textrm{B} B$ is the Zeeman splitting,
$E_\alpha=\alpha \sqrt{2eB/\hbar}$ with $\alpha$ the Rashba coupling constant.
The $N=0$ state is fully spin-polarized, and the two series of LLs with $N>0$
correspond to orthogonal mixtures of spin-up and spin-down states.

To compare the data with this second model, we computed numerically the DOS, the
chemical potential, and the conductance for each magnetic field and temperature,
using the formalism of Ref.~\onlinecite{Ando1974}. We considered a Gaussian
broadening of the LLs with a variance $\gamma^\pm \sqrt{B}$. 
The results are displayed in Fig.~\ref{RZmodel}. As can be seen, good agreement
between the data and the theory is obtained (see supplementary data).

\begin{figure*}
\setlength\fboxsep{0pt}
\setlength\fboxrule{0.25pt}
\centering\includegraphics[width=6in]{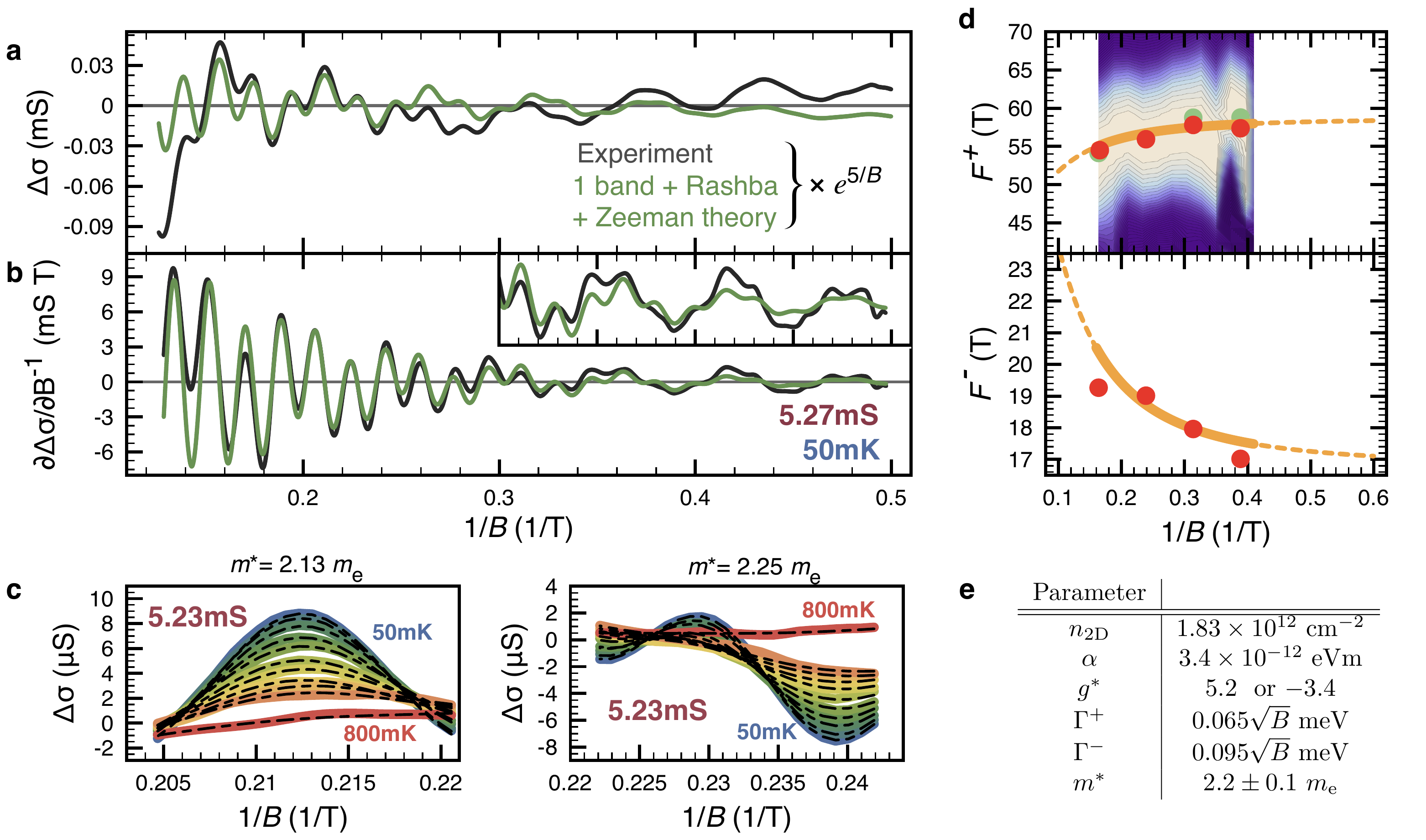}
\caption{\label{RZmodel}
(a) Comparison between $\Delta\sigma$ measured experimentally for
$\sigma_{\textrm{2D}}=5.27$~mS (black) and calculated using our single-band
model with Rashba and Zeeman interactions (green). (b) Derivative with
respect to $B^{-1}$ of the curves presented in (a). The exponential factor $e^{5/B}$ is used to magnify the low-field
region. (inset) Close-up on the low-field region of (b). (c) Temperature dependence of the oscillations over
the two ranges of applied field used to extract the effective mass. Black dashed
lines are the theoretical curves computed using $m^*\sim 2.2 m_e$, while the
thick coloured lines are the experimental data. (d) Magnetic field  dependence of $F^{+}$ (top) and $F^{-}$ (bottom). Red and green dots correspond, respectively, to the estimation made via the band pass filtered and differentiated signals. The background is a color plot based on the short time Fourier Transform of $\Delta \sigma$. (e) Summary of the fitting parameters.}
\end{figure*} 
The carrier density extracted from this analysis is
$n_{\textrm{2D}}^{\textrm{SdH}}=\cdensityunit{1.83}$, lower than
the one found using the Hall effect measurements. The magnitude of the obtained Rashba coupling
constant ($\alpha=\alphaunit{3.4}$) agrees very well with values obtained from
weak localization analyses and from modelling of the transport data in parallel
fields \cite{CavigliaWAL, BenShalom, AlexAMRO}. We note that, given the small value of $k_{\textrm{F}}$ in our samples, a $k$-cubic Rashba interaction inducing a spin-splitting of $\approx \SI{2}{\milli \electronvolt}$ would require a very large coupling constant, beyond values recently reported \cite{Nakamura2012}.

To obtain the effective mass, we selected three peaks from a region of magnetic
fields where the amplitude of the oscillations is large. Fig.~\ref{RZmodel}c
shows that the data can be fit perfectly using an effective mass of $2.2\pm
0.1 m_e$. This value may indicate that the electronic state of the
oscillating carriers is not dominated by Ti $d_{xy}$ orbitals, as one
would then expect a lower effective mass ($\lesssim m_e$). Instead, the higher mass
obtained in this analysis can be understood by taking into account the
contribution of $d_{xz}$/$d_{yz}$ orbitals to the electronic states. We note that a recent analysis of photoemission spectra for interfaces grown at \SI{650}{\celsius}, complemented by \textit{ab initio} calculations, was consistent with a 2DEG having occupied $d_{xz}$/$d_{yz}$ electronic states at the Fermi energy \cite{Cancellieri2014}. This observation corroborates our recent results on standard LAO/STO interfaces,
where a sharp decrease in the elastic scattering rate was correlated to the
progressive appearance, at the Fermi level, of heavier carriers \cite{AlexAMRO}.

A lingering question pertains to the explanation of the lower carrier density and high mobility that are measured in samples prepared at low growth temperature, as compared to the ``standard'' ones discussed in the introduction. A puzzle related to this issue concerns  the systematic discrepancy in the value of the carrier concentration that one finds when comparing Hall and SdH data \cite{CavigliaSdH,BenShalom2010}. One may surmize that these observations point to the critical role played in transport by the two different types of electronic orbitals of the $t_{2g}$ triplet. Spectroscopies and DFT calculations show that $d_{xy}$ states are located close to the interface where disorder and lattice distorsions likely result in low mobility \cite{Delugas}. Heavy $d_{xz}$/$d_{yz}$ sub-bands extend deeper into the STO bulk and hence are less sensitive to these effects, giving rise to a much higher mobility; the presence of a large Rashba splitting could further help explain an enhancement of  this mobility due to protection against backscattering. While both types of orbital can contribute to the magnetoresistance (analysis of our high-mobility samples support that), only the heavy $d_{xz}$/$d_{yz}$ states have a high enough mobility to sustain SdH oscillations in our accessible range of magnetic fields.

Because the Zeeman energy enters equation~\Ref{RZLLs} only as a squared term,
for the LLs with $N>0$, we find two solutions for the $g^*$-factor, namely $5.2$
or $-3.4$,  values similar to the ones observed in semiconductor heterostructures. We note that $g^*$-factors significantly different from 2 were predicted by \textit{ab initio} calculations in bulk STO \cite{VanderMarel2011}. In this second scenario, we can also estimate the Rashba splitting and the Fermi energy. Interestingly, we find that both are of the same order
of magnitude: $\Delta_R=2.2$~meV and $E_{\textrm{F}}=1.65$~meV, a situation  very
different from the one of many semiconductor 2DEGs, where the Fermi
energy dominates. 

Owing to the complexity of the Rashba LLs spectrum, the oscillation frequencies ($F^-$, $F^+$ now linked to the $-$ and $+$ levels) are predicted to be field dependent \cite{Novokshonov2006,Islam2012}. We have estimated, from the data, $F^-$ and $F^+$ as a function of magnetic field using three different procedures (see supplementary data). Fig.~\ref{RZmodel}d shows the estimated $F^{-}$ and $F^{+}$ as a function of $1/B$ on top of the theoretical prediction (thick orange line). A very good agreement is obtained both in the amplitudes and in the signs of the frequency variations. We surmize that the field dependence of $F^{-}$ and $F^{+}$ is the reason why the low field region of the quantum oscillations is fitted better by the Rashba model than by the two-band model (compare plots in inset of Fig.~\ref{MBmodel}b and Fig.~\ref{RZmodel}b). The fact that the pseudo-frequencies $F^{-}$ and $F^{+}$  depend on $B$ is pointing to SdH oscillations originating from a Rashba spin-split band and not from two bands.

\begin{figure*}
\centering\includegraphics[width=2\columnwidth]{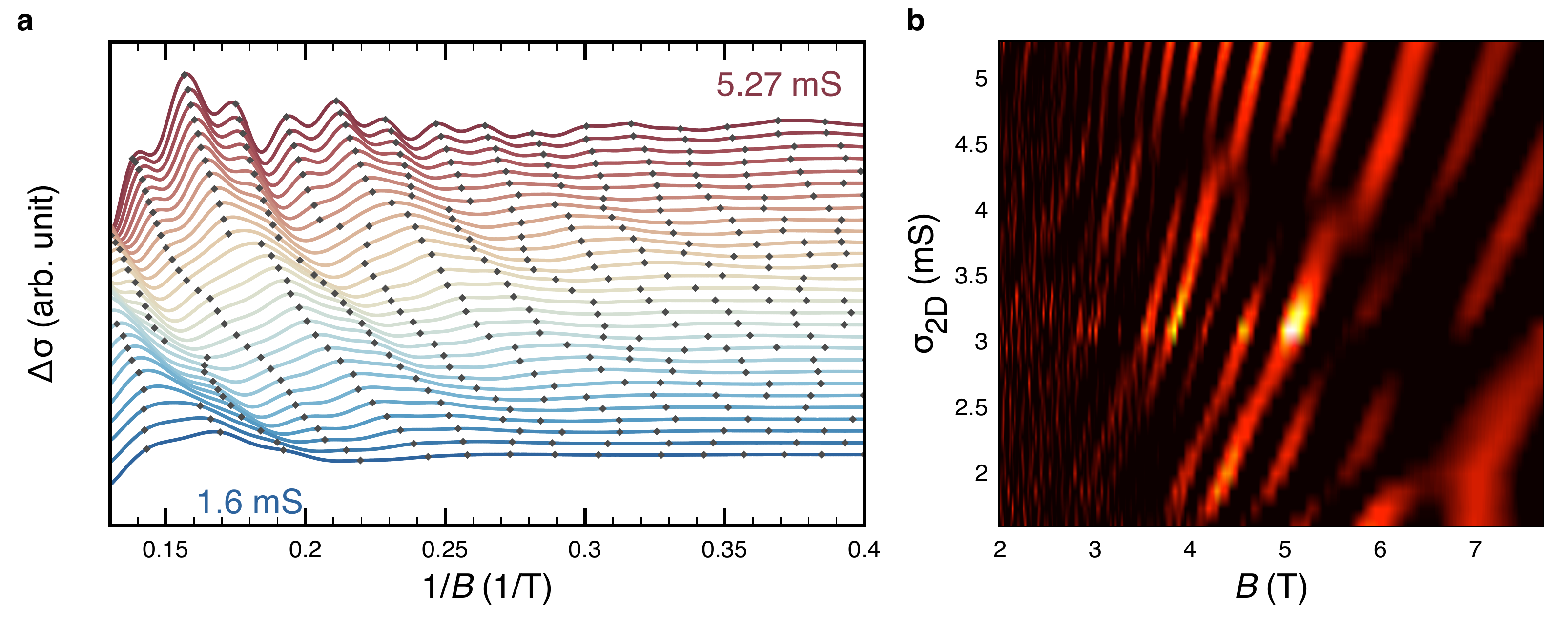}
\caption{\label{Fig4}
(a) Analysis of the doping dependence of the SdH oscillations. Curves are shifted for clarity. Black dots
indicate the position where maxima occur in $-\Delta\sigma''(1/B)$. (b) Fan diagram showing $-\Delta\sigma''(1/B)$; yellow corresponds to the maxima
and black to negative values.}
\end{figure*}

We finally discuss the gate-voltage dependence of the SdH oscillations.
Fig.~\ref{Fig4}a shows the change in conductance as a function of $1/B$ for
various $V_g$ (\textit{i.e.} $\sigma_\textrm{2D}$). A clear evolution of the SdH
oscillations with decreasing doping is visible and is compatible with the shrinking of the
Fermi surface expected from Hall measurements. 
With the help of the second derivative $-\Delta\sigma''(1/B)=-\partial^2 \Delta \sigma/\partial (1/B)^2$ which amplifies $F^{+}$, we identify all maxima from the ``$+$'' levels as a function of applied applied gate voltage. These maxima are indicated in Fig.~\ref{Fig4}a by black dots. We expect that the
trajectories traced out by the black dots as a function of $V_g$ correspond to
the evolution of each LL as a function of the chemical potential. Strikingly, we
see that these trajectories present sharp deviations or jumps upon decreasing
$V_g$.
This feature is clearly visible on the fan diagram of Fig.~\ref{Fig4}b featuring $-\Delta \sigma''(1/B)$ versus $(B, \sigma_\textrm{2D})$  which nicely illustrates the fact that the position of the ``$+$'' LLs follows a
simple evolution only for limited regions of the diagram. Conversely, we observe that at precise locations the amplitude of the SdH oscillations is strongly suppressed.

There are many situations in which quantum oscillations rapidly change their phase and/or amplitude as a function of $B$. An example is the exchange interaction that enhances the $g^*$-factor for magnetic fields beyond a critical value \cite{Janak1969,Nicholas1988,Fogler1995,Leadley1998,Brosig2000,Piot2005,Yang2008,Krishtopenko2012}. Changes in the oscillation pattern can also occur when different LLs cross at a particular magnetic field: in this case anti-crossings can be observed \cite{Muraki2001,Zhang2005, Zhang2006, Ellenberger2006, Yu2007,Jungwirth2000,Smith2012}. These phenomena originate from many-body interactions. The deviations observed in Fig.~\ref{Fig4} point to an interaction whose energy scale is of the order of the LL splitting ($\approx \SI{0.1}{\milli \electronvolt}$ at \SI{2.5}{\tesla} for the highest doping and the ``+'' levels). Further studies are needed to determine the nature of this interaction.

The study presented here unravels the remarkably complex behavior of the
Shubnikov-de Haas oscillations seen at the LAO/STO interface. Our analysis reveals the important
role played by the Rashba SOi on the electronic band structure and the peculiar regime hereby realized. Finally, the evolution of
the LL spectrum as a function of doping and magnetic field displays sharp deviations that we cannot explain in our independent electron picture.

The authors would like to thank G.~Seyfarth and D.~Jaccard for help with the
measurements and stimulating discussions, and are grateful to M.~Lopes and
S.~C.~M{\"u}ller for their technical assistance. This work was supported by the
Swiss National Science Foundation through the NCCR MaNEP and Division II, by
the Institut Universitaire de France (MG) and has received funding from the European Research Council under the European Union's Seventh Framework Programme (FP7/2007-2013) / ERC Grant Agreement n$^\circ$ 319286 (Q-MAC).

\clearpage

\section{\boldmath L\lowercase{arge modulation of the }S\lowercase{hubnikov-de }H\lowercase{aas oscillations by the }R\lowercase{ashba interaction at the }L\lowercase{a}A\lowercase{l}O$_{3}$/S\lowercase{r}T\lowercase{i}O$_{3}$\lowercase{ interface - }S\lowercase{upplemental }M\lowercase{aterial}}

\setcounter{figure}{0}

\subsection{S1 -- Growth conditions and sample geometry}

LaAlO$_3$/SrTiO$_3$ interfaces were realized by growing 9~unit cells of LaAlO$_3$ on a
(001) oriented TiO$_2$ terminated SrTiO$_3$ substrate using pulsed laser
deposition. The deposition conditions were an oxygen pressure of
$10^{-4}$~mbar, a substrate temperature of 650~$^{\circ}$C, a
repetition rate of the ablating laser of 1~Hz, and a fluence of \SI{0.6}{\joule \per \square \centi \meter}.
The growth process was monitored \textit{in-situ} using reflection high energy
electron diffraction (RHEED). After growth, the sample was annealed for 1~hour
in 0.2~bar of O$_2$ at a temperature of $\sim 530^{\circ}$C.

\begin{figure}[!h]
\centering\includegraphics[width=\columnwidth]{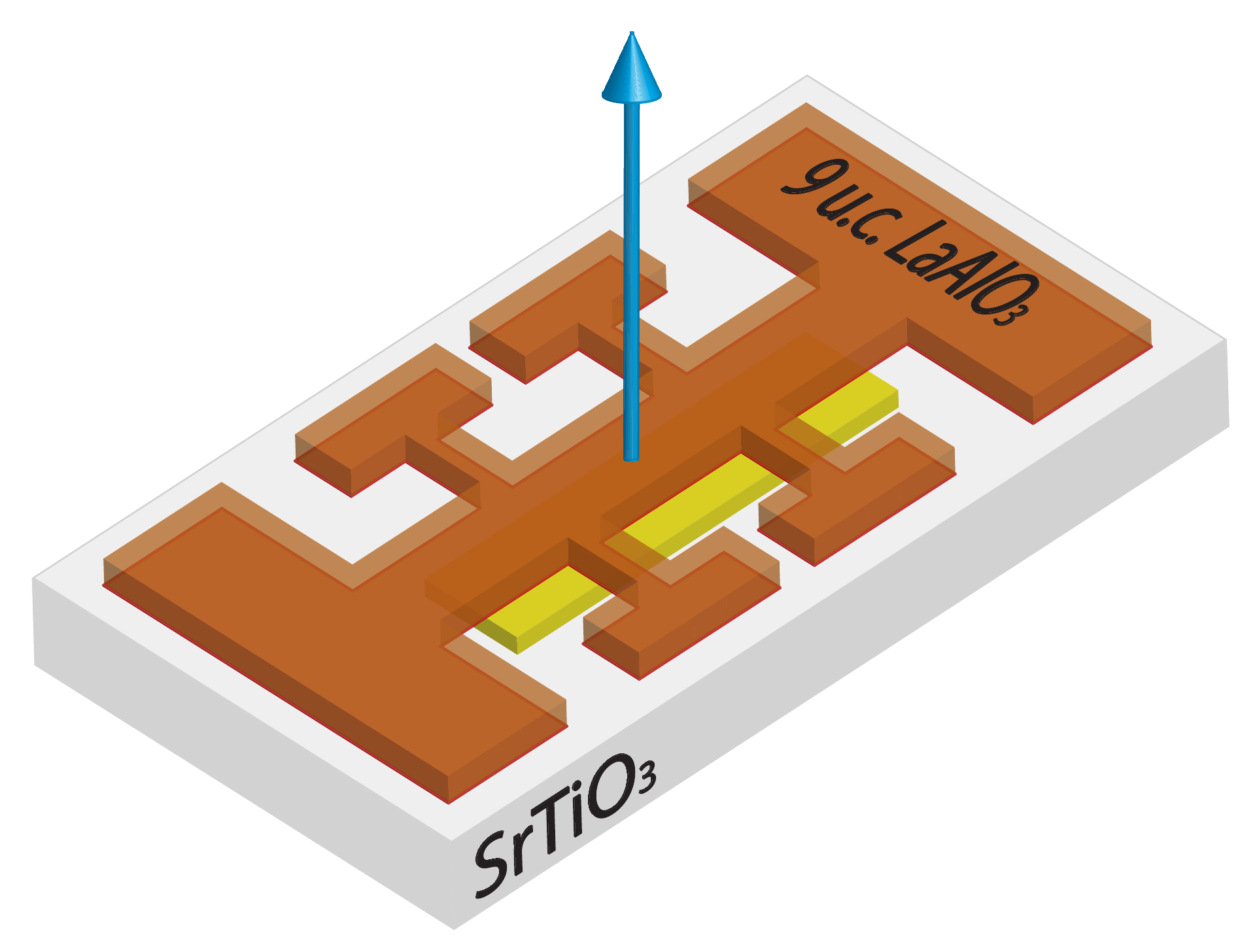}
\caption{\label{Device}
Schematic view of sample geometry.
\label{samplegeom}
}
\end{figure}

In order to avoid any photolithographic step after the layer deposition, we
patterned the substrate with amorphous SrTiO$_3$ \cite{Stornaiuolo2012}. The
dimensions of the Hall bars for magnetotransport measurements were $500 \mu
\textrm{m}\times 1000 \mu\textrm{m}$ (width$\times$length). The field-effect
devices were realized using the STO substrate as the gate dielectric adding a metallic contact on its backside (yellow rectangle, see Fig.~\ref{samplegeom}). The blue arrow indicates the direction of the external magnetic field for all the magnetotransport measurements of this work.

\subsection{S2 -- Effective mass associated with the high- and low-frequency components of the SdH oscillations}

In the LK formalism, the temperature evolution of the oscillations has the
following functional form:
	\begin{equation}\label{Tdep}
		\Delta\sigma_i(B_{m},T)\propto\frac{\alpha_{\textrm{LK},i}T}
		{\sinh(\alpha_{\textrm{LK},i} T)}
	\end{equation}
with $\alpha_{\textrm{LK},i}=2\pi^2 k_\textrm{B}/\hbar\omega_{c,i}^*$,
$\omega_{c,i}^*=e B_{m}/m_i^*$, $B_{m}$ the field at which the extremum is
observed, $i$ the band index, $T$ the temperature, and $-e$ the electronic
charge.

\begin{figure}[t]
\centering\includegraphics[width=0.4\textwidth]{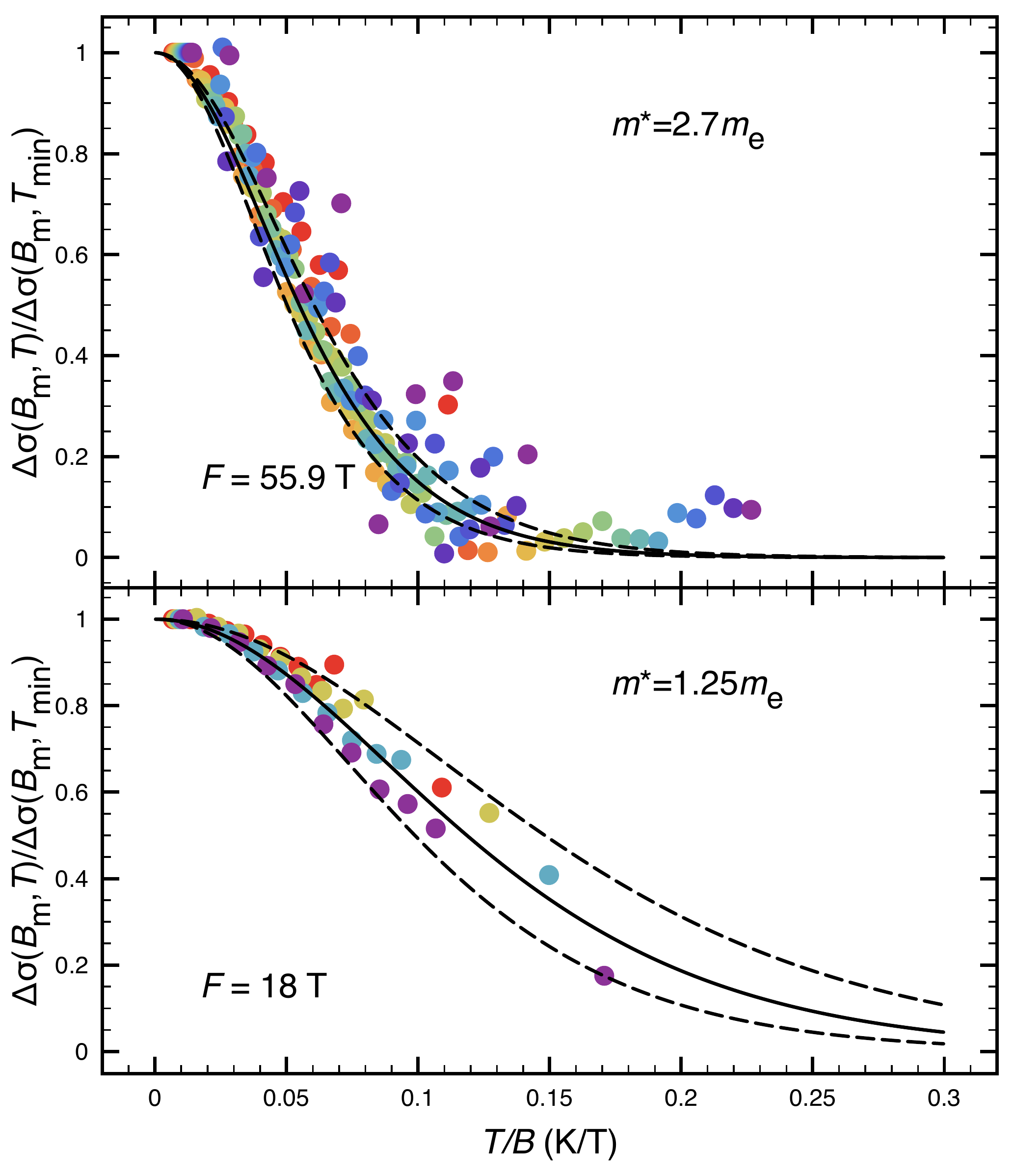}
\caption{\label{EffMass}
Extraction of the effective mass from the temperature dependence of the Shubnikov-de Haas oscillations in the two band model. Full lines correspond to the theoretical predictions taking $m^*$ as effective mass. Dashed lines correspond to the theoretical predictions taking $m^*\pm0.25 m_\textrm{e}$.
}
\end{figure}

Figure~\ref{EffMass} illustrates the procedure that we followed to extract the effective mass using Eq.~(\ref{Tdep}). The top (bottom) panel shows data obtained for the high-frequency (low-frequency) component of the magnetoconductance recorded at $\sigma_{\textrm{2D}}=5.27$~mS. Each color is linked to the temperature evolution of a single oscillation.

\subsection{S3 -- Modeling the SdH oscillations in the case of a single band with Rashba/Zeeman splitting}

\begin{figure}[!h]
\centering\includegraphics[width=\columnwidth]{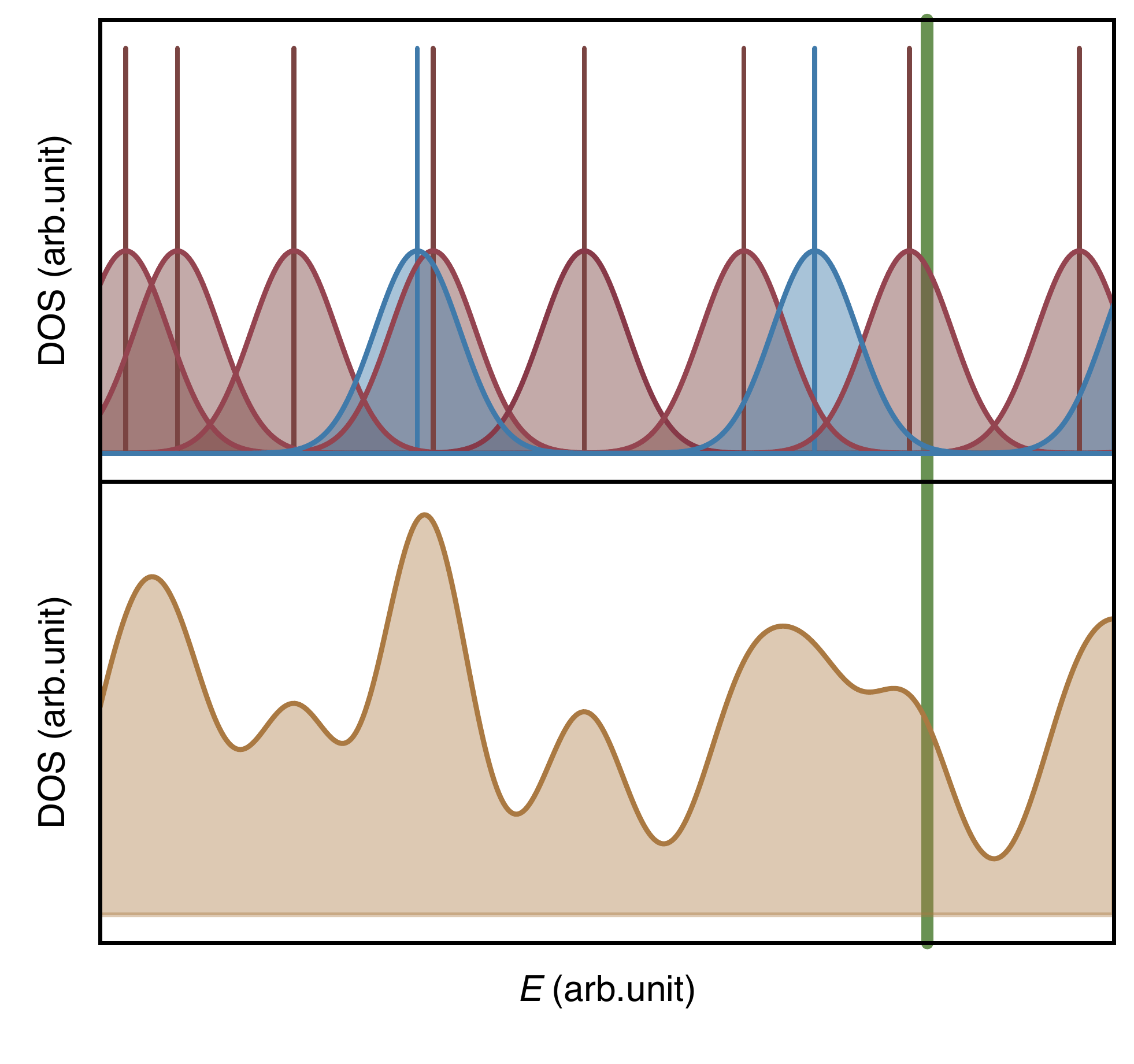}
\caption{\label{IndividualDOS}
(Top) Schematic view of the DOS for the two families ($+$, red; $-$, blue) of LLs in the Rashba model. (Bottom) Total DOS resulting from the sum of the $+$ and $-$ DOS. The green vertical line is the position of the Fermi level.
}
\end{figure}

In order to determine the SdH oscillations pattern of the conductance for a single band with Rashba and Zeeman splittings, we start with the expression of the Landau levels presented in the main text,
	\begin{equation}\label{RZLLs}\begin{split}
		E_{N=0} & =E_c/2-E_Z \\
		E_{N>0}^\pm & =N E_c\mp\sqrt{(E_c/2-E_Z)^2+N E_\alpha^2}\,,
	\end{split}\end{equation}
with $E_c=\hbar\omega_c^{*}$, $E_Z=(g^*/2)\mu_\textrm{B} B$ the Zeeman splitting,
and $E_\alpha=\alpha\sqrt{2eB/\hbar}$ with $\alpha$ the Rashba coupling. Setting the values of $m^*$, $g^*$, $\alpha$, and $B$ defines the energy of the Landau levels (LLs) for a given magnetic field (vertical red/blue lines in Fig.~\ref{IndividualDOS}, top). We broaden each level using a gaussian line shape with a variance $\Gamma^{\pm}=\gamma^{\pm}\sqrt{B}$, and perform the sum over all levels to obtain the density of states $g(E)$:
	\begin{equation}\label{DOS}
		g(E)=\frac{eB}{2\pi\hbar}\sum_{N,\,s=\pm}
		\frac{1}{\sqrt{2\pi}\Gamma^s}
		\exp\left[-\frac{1}{2}\left(\frac{E-E_N^s}{\Gamma^s}\right)^2\right].
	\end{equation}
The result is shown in Fig.~\ref{IndividualDOS}, bottom.

Different schemes are possible for the Landau-level broadening. In addition to the Gaussian with constant variance, Lorentzian and semi-elliptic line shapes have been considered \cite{Usher2009}. In certain circumstances, a constant background was added to the total DOS. However, according to Ref.~\onlinecite{Zhou2012}, many experiments revealed either a Gaussian line shape with a variance proportional to $\sqrt{B}$, or a Lorentzian one.

The Gaussian line shape is easier to treat numerically, thanks to the rapid decay of the Gaussian and the $\sqrt{B}$ dependence of the variance allows a better fitting of our data.

The chemical potential $\mu$ (vertical green line in Fig.~\ref{IndividualDOS})   is obtained by solving numerically the equation giving the carrier concentration at a given temperature,
	\begin{equation}\label{mudetermination}
		n_{\textrm{2D}}=\int_{-\infty}^{\infty}dE\,f(E-\mu)g(E),
	\end{equation}
with $f(E)=(e^{E/k_{\textrm{B}}T}+1)^{-1}$ the Fermi-Dirac distribution. Finally, the conductance is obtained via \cite{Ando1974}:
	\begin{multline}\label{MCdetermination}
		\sigma_{xx}=\frac{e^2}{2\pi\hbar}\int_{-\infty}^{\infty}dE\,
		\left(-\frac{\partial f(E-\mu)}{\partial E}\right) \\
		\times\sum_{N,\,s=\pm}\left(N+\frac{1}{2}\right)
		\exp\left[-\left(\frac{E-E_N^s }{\Gamma^s}\right)^2\right].
	\end{multline}
We performed this calculation at each $B$ in order to get the field dependence $\sigma_{xx}(B)$. Finally, the oscillating part $\Delta\sigma_{xx}(B)$ of the conductance was obtained by subtracting a curve computed using a sufficiently high temperature $T_{\textrm{high}}$, at which the SdH oscillations are completely suppressed:
	\begin{equation}
		\Delta\sigma_{xx}(B,T)=\sigma_{xx}(B,T)-\sigma_{xx}(B,T_{\textrm{high}}).
	\end{equation}
This method was successfully applied, e.g., in Ref.~\onlinecite{Luo1990}.

\subsection{S4 -- Minimal parametrization of the Landau-level spectrum in the case of a Rashba/Zeeman split system}

In the presence of a linear Rashba spin-orbit interaction and a Zeeman splitting, the LLs spectrum of a parabolic and isotropic band is given by Eq.~(\ref{RZLLs}). Introducing the energy scale $E_{c,1}\doteqdot E_{c}(B=1~\textrm{T})$ to normalize the energies, $\tilde{E}=E/E_{c,1}$, Eq.~(\ref{RZLLs}) can be simplified:
	\begin{equation}\label{RZLL0newunit}
		\tilde{E}_0 =\frac{E_c}{2E_{c,1}}-\frac{E_Z}{E_{c,1}}
		=\frac{B}{2}\left(1-\frac{g^*m^*}{2m_e}\right) \doteqdot \frac{Ba}{2}
	\end{equation}
with $a=1-g^*m^*/(2m_e)$. Similarly,
	\begin{align}\label{RZLLsNnewunit}
		\nonumber
		\tilde{E}_N^\pm &= N\frac{E_c}{E_{c,1}}
		\mp\sqrt{\left(\frac{E_c}{2E_{c,1}}-\frac{E_Z}{E_{c,1}}\right)^2
		+N\frac{E_\alpha^2}{E_{c,1}^2}} \\
		\nonumber
		&=NB\mp\sqrt{\left(\frac{B a}{2}\right)^2
		+N\frac{2(\alpha m^*)^2}{e\hbar^3}\frac{1}{B}} \\
		&= B\left(N\mp\frac{1}{2}\sqrt{a^2+N\frac{D}{B}}\right)
	\end{align}
with $D=8(\alpha m^*)^2/(e\hbar^3)$.
We observe that, in the constant $E_{\textrm{F}}$ approximation, only three parameters are required in order to determine the values $B^*$ where the conductance is maximal \footnote{Maxima in the conductance arise when a Landau level is at the Fermi energy, i.e., when $B$ is such that one of the equations (\ref{RZLL0newunit}) or (\ref{RZLLsNnewunit}) is verified with $\tilde{E}=\tilde{E}_{\textrm{F}}$.}: those are $\tilde{E}_{\textrm{F}}$, $a$ and $D$.

\subsection{S5 -- Using the LK formula to analyze the quantum oscillations in Rashba spin-orbit split bands}

In this section we show that the LL spectrum generated by the Rashba spin-orbit interaction gives rise to SdH oscillations with two ``pseudo-frequencies'' that are field dependent.

It is well known that the LLs spectrum generated by a parabolic band is :
	\begin{equation}\label{LLevels}
		E_n=\hbar \omega_c^*\left(n+\frac{1}{2}\right)
		=\hbar \frac{eB}{m^*}\left(n+\frac{1}{2}\right)
	\end{equation}
Hence, in this model, the energy splitting between neighboring Landau levels is directly linked to the value of the effective mass and is independent of $n$. Indeed:
	\begin{equation}
		\frac{d E_n}{d n}=\hbar \frac{eB}{m^*}
		\label{LLevelssplittingsimple}
	\end{equation}
For a Rashba spin-split band the use of Eq.~(\ref{RZLLsNnewunit}) leads to:
	\begin{equation}\label{LLevelssplittingfermilevelRashba}
		\frac{d E_{N>0}^\pm}{d N}=\hbar \frac{eB}{m^*} \left(1\mp \frac{D}{4 B}\frac{1}{\sqrt{a^2+N\frac{D}{B}}}\right)
	\end{equation}
In this case, the splitting between LLs is $N$--dependent. 

As we consider the conductance which is related to the DOS at the Fermi level, this spacing can be considered as almost constant for large $N$. More generally, we calculate, at a given magnetic field, the LL index at the Fermi level ($N^{\pm}_\textrm{F}$). Using the constant $E_\textrm{F}$ approximation and restricting ourselves to $E_\textrm{F} \ge 0$ we get:
 	\begin{equation}
		\label{NvsB}
		N^{\pm}_\textrm{F}=\frac{m^* E_\textrm{F}}{\hbar e B}+\frac{D}{8 B}\pm \frac{\kappa}{8}
	\end{equation}
with:
 	\begin{equation}
		\kappa=\sqrt{16 a^2+\frac{D}{B^2}\left(D+16 \frac{m^* E_\textrm{F}}{\hbar e}\right)}
	\end{equation}
In turn, Eq.~(\ref{NvsB}) can be used to define two ``pseudo-frequencies'' via:
	\begin{equation}
	F^\pm(1/B)=\frac{d N^{\pm}_\textrm{F}}{d (1/B)}
	\end{equation}
We get:
	\begin{equation}
	F^\pm(1/B)=\frac{m^* E_\textrm{F}}{e \hbar} + \frac{D}{8}\pm \left( \frac{\kappa}{8}-\frac{2 a^2}{\kappa}\right)B
	\end{equation}
We observe that $F^+(1/B)$ and $F^-(1/B)$ depend on the magnetic field strength; this is the reason why we call them ``pseudo-frequencies''. This dependence is due to the non-linear spacing of the Rashba/Zeeman split LLs. $F^+(1/B)$ is a decreasing function of $B$ while $F^-(1/B)$ is a increasing function of $B$. As shown in the main text, even though this magnetic field dependence is weak, it can be evidenced in our experimental oscillations.

For the experimental determination of $F^{+}$ as a function of $1/B$ we used two techniques. The first one consists in using filtering techniques or the second derivative of $\Delta \sigma$ to isolate the oscillations associated to $F^{+}$ (red and green dots in Fig.~3d of the main text), dividing the field range in 4 regions of equal size in $1/B$ and fitting the oscillations (in each region) with a cosine of constant frequency. The second one is to compute the short time Fourier transform (FT) of $\Delta \sigma$ (color plot in Fig.~4a of the main text). For $F^{-}$, due to the limited number of oscillations, we could only apply the first technique.

Coming back to Eq.~(\ref{LLevelssplittingfermilevelRashba}), the splitting between LLs at the Fermi level in a Rashba/Zeeman scenario is obtained by inserting Eq.~(\ref{NvsB}) into Eq.~(\ref{LLevelssplittingfermilevelRashba}):
	\begin{equation}
		\left(\frac{d E_{N>0}^\pm}{d N}\right)_{E=E_\textrm{F}}=\hbar \frac{eB}{m^*} \left(1\mp D\frac{1}{B \kappa \pm D}\right)
	\end{equation}	
which, if we define
	\begin{equation}
		m^*_{\pm}=m^*\left( 1 \pm \frac{D}{B \kappa} \right)
	\end{equation}	
can be rewritten in the same form as Eq.~(\ref{LLevelssplittingsimple}):
	\begin{equation}
		\left(\frac{d E_{N>0}^\pm}{d N}\right)_{E=E_\textrm{F}}=\hbar \frac{eB}{m^*_{\pm}} 
	\end{equation}
Hence, in a magnetotransport experiment, if $m^*_{\pm}$ does not vary too much with magnetic field, a Rashba/Zeeman split LLs spectrum can be interpreted as two independent series of LLs of the form given by Eq.~(\ref{LLevels}) (\textit{i.e.} an analysis of the quantum oscillations using the LK formula will not fail altogether). However, in this case, the effective mass associated with $F^+(1/B)$ is larger than the effective mass associated with $F^-(1/B)$. This is exactly what we find.
	
We note that, with the parameters extracted from the fit shown in Fig. 3 of the main text, we find that the dependence of $m^*_{\pm}$ on magnetic field is only of 5--10\% between $2$ and \SI{8}{\tesla}, which is below our experimental resolution.

Formula similar to the ones found in this section can be found in Refs.~\onlinecite{Novokshonov2006,Islam2012} .

\subsection{S6 -- Determination of the effective masses associated with the quantum oscillations stemming from inter-sub-band scattering}
In a 2DEG with more than one sub-band populated, inter-sub-band scattering was shown to bring additional components to the Shubnikov-de Haas oscillations \cite{Coleridge1990,Leadley1992}. More recently, this effect was also discussed for systems composed of a unique Rashba spin-orbit split band \cite{Novokshonov2013}: the (pseudo-)frequencies associated to this phenomenon are $F^{+}+F^{-}$ and $F^{+}-F^{-}$. Moreover, according to theory, the two components related to the inter-sub-band scattering have a different temperature behavior; a higher effective mass is associated to the one at $F^{+}+F^{-}$ than to the one at $F^{+}-F^{-}$.

\begin{figure}
\centering\includegraphics[width=\columnwidth]{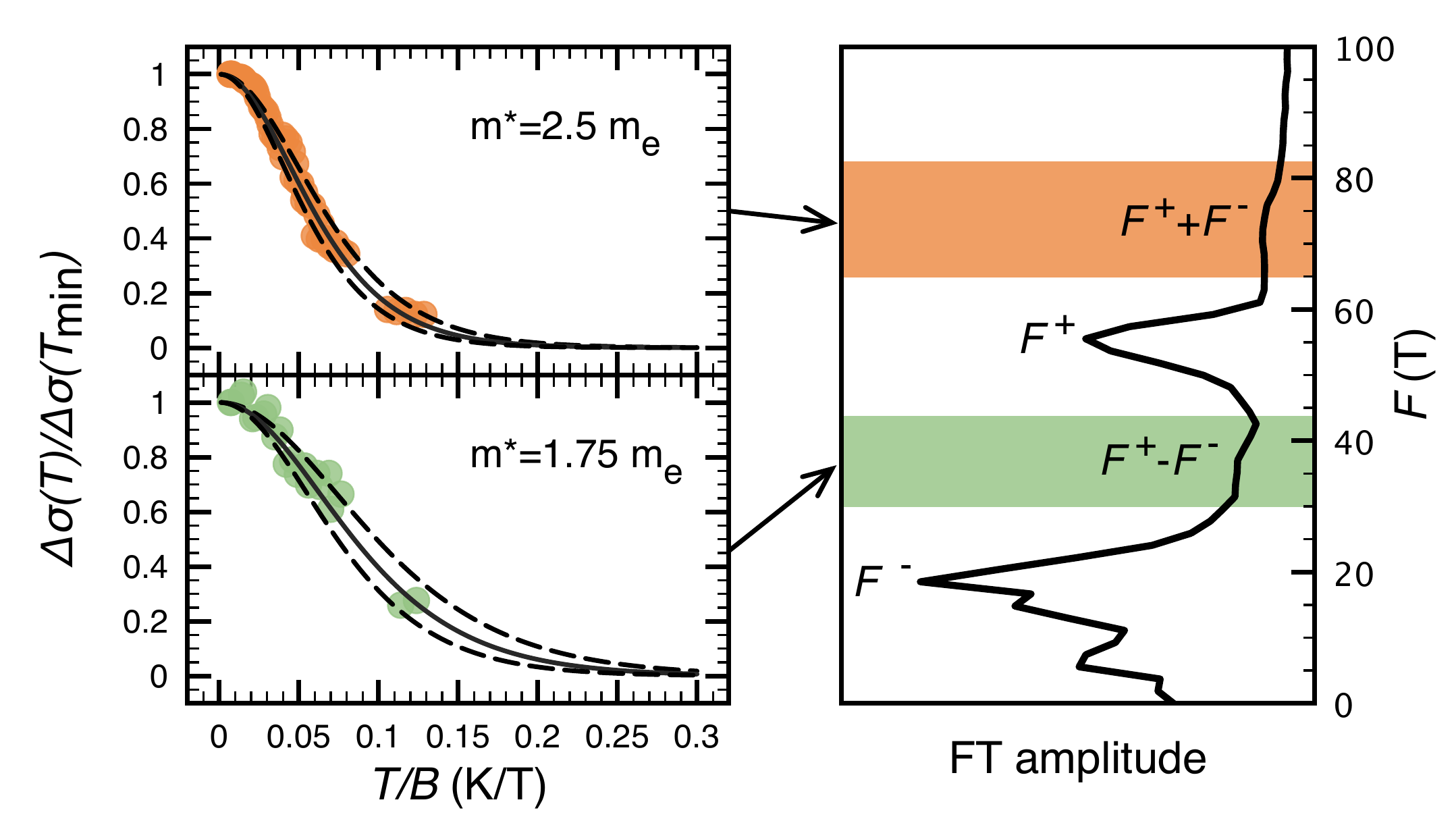}
\caption{(Left) Temperature dependence of the FT components located at $F^{+}-F^{-}$ and $F^{+}+F^{-}$, for the measurement at the doping with the highest conductance (5.27 mS). Full lines correspond to the theoretical predictions taking $m^*$ as effective mass. Dashed lines correspond to the theoretical predictions taking $m^*\pm0.25 m_\textrm{e}$. (Right) Fourier transform of the same measurement. Colored regions define the limits of the band-pass filters.}
\label{EffMass_intersubbandpeaks}
\end{figure}

In Fig. \ref{EffMass_intersubbandpeaks}, we present an analysis of the temperature dependence of the oscillations associated with the maxima of the FT that we indeed can observe at $F^{+}+F^{-}$ and $F^{+}-F^{-}$, for the measurement at the doping with the highest conductance (5.27 mS). We used band pass filters (whose limits are illustrated in Fig. \ref{EffMass_intersubbandpeaks}) to select the different components of our spectrum. We obtain an effective mass of $1.75 m_\textrm{e}$ and $2.5 m_\textrm{e}$ for the component at $F^{+}-F^{-}$ and $F^{+}+F^{-}$, respectively. Due to the weak contribution of these frequencies to the total FT a non-negligible weight from the neighboring peaks is probably biasing our estimations.

Using the formula in \cite{Novokshonov2013}, and the parameters of the fit presented in Fig.~3 of the main text, we calculate a mass of $\approx1.2m_\textrm{e}$ (at $B=\SI{5}{\tesla}$) for the peak at $F^{+}-F^{-}$, in reasonable agreement with our result.

\end{document}